\documentclass[a4paper,twoside]{article}
\usepackage{subcaption}
\usepackage{calc}
\usepackage{amssymb}
\usepackage{amstext}
\usepackage{amsmath}
\usepackage{amsthm}
\usepackage{multicol}
\usepackage{pslatex}
\usepackage{apalike}
\usepackage{algorithm2e}
\usepackage[bottom]{footmisc}
\usepackage{listings}
\usepackage{inconsolata}
\usepackage{xcolor}
\usepackage{physics}
\usepackage{fancyvrb}
\usepackage{dsfont}
\usepackage[hidelinks]{hyperref}
\VerbatimFootnotes
\usepackage{SCITEPRESS}

\newcommand{\codefontsize}{\fontsize{7.5}{9.5}\selectfont}
\lstdefinestyle{mintedlike}{
  language=Python,
  basicstyle=\codefontsize\ttfamily,
  keywordstyle=\color{blue!90!black}\bfseries,
  commentstyle=\color{gray!70!black}\itshape,
  stringstyle=\color{red!70!black},
  numberstyle=\tiny\color{gray},
  numbers=left,
  numbersep=8pt,
  stepnumber=1,
  showstringspaces=false,
  breaklines=false,
  breakatwhitespace=true,
  rulecolor=\color{gray!50},
  tabsize=2,
}
\begin{document}

\title{\texttt{randextract}: a Reference Library to Test and Validate\\Privacy Amplification Implementations}

\author{
\authorname{Iyán Méndez Veiga\sup{1,2}\orcidAuthor{0009-0004-8249-4661}, Esther Hänggi\sup{1}\orcidAuthor{0000-0002-9760-8535}}
\affiliation{\sup{1}School of Computer Science and Information Technology, Lucerne University of Applied Sciences and Arts,\\6343 Rotkreuz, Switzerland}
\affiliation{\sup{2}Institute for Theoretical Physics, ETH Zurich, 8093 Zurich, Switzerland}
\email{iyan.mendezveiga@hslu.ch}
}


\abstract{Quantum cryptographic protocols do not rely \emph{only} on quantum-physical resources, they also require reliable classical communication and computation. In particular, the secrecy of any quantum key distribution protocol critically depends on the correct execution of the \emph{privacy amplification} step. This is a classical post-processing procedure transforming a partially secret bit string, known to be somewhat correlated with an adversary, into a shorter bit string that is close to uniform and independent of the adversary's knowledge. It is typically implemented using randomness extractors. Standardization efforts in quantum cryptography have focused on the security of physical devices and quantum operations. Future efforts should also consider all algorithms used in classical post-processing, especially in privacy amplification, due to its critical role in ensuring the final security of the key. We present \texttt{randextract}, a reference library to test and validate privacy amplification implementations.}

\onecolumn \maketitle \normalsize \setcounter{footnote}{0} \vfill

%%
%% 1. INTRODUCTION
%%
\section{\uppercase{Introduction}}\label{sec:introduction}

Quantum cryptography is a multidisciplinary field taking advantage of quantum features, such as entanglement or the impossibility of cloning quantum states, to design cryptographic protocols. Two such protocols are Quantum Key Distribution (QKD)~\cite{bb84,ekert} and Quantum Random Number Generators (QRNGs)~\cite{oqrng}. QKD allows two honest parties to establish a secret key over an insecure public quantum channel, and QRNGs can produce truly unpredictable randomness.

Besides the physical implementation using quantum technology, quantum cryptographic protocols rely on classical computation and communication. In particular, both QKD and QRNGs depend on classical data processing steps to achieve their goals. One of these steps, privacy amplification~\cite{privacy-amplification}, is crucial for security. Bugs or deviations from a correct implementation can introduce vulnerabilities and compromise the security guarantees of the protocols. In the case of QKD, this could result in the generation of insecure keys.

Current standardization efforts in quantum cryptography have focused on the security of physical devices and quantum operations~\cite{ISO23837-2:2023,ETSI-GS-QKD011:2016}. To ensure the security of the complete protocol, these efforts have to be extended to include all classical post-processing tasks. This includes the careful selection and standardization of algorithms, rigorous testing against publicly available test vectors, and validation through well-established certification programs. In contrast to classical cryptographic algorithms and devices~\cite{NIST-CMVP:2025,NIST-CAVP:2025}, such processes do not yet exist for quantum cryptography.

\subsection{Our Contribution}

Our work provides a reference implementation of the functions used in privacy amplification allowing to check for correctness.

\begin{enumerate}
    \item \texttt{randextract} is an open-source \textbf{Python library} that implements (modified) Toeplitz hashing and Trevisan’s extractor, two types of functions commonly used in the privacy amplification step. It is distributed as an easy-to-use Python package. This allows the source code to remain close to the mathematical formulations, reduces the possibility of implementation bugs due to peculiarities of the programming language and improves readability, allowing a large audience to read and audit the code. The library focuses on correctness, containing hundreds of unit and integration tests, and is not performance-optimized.

    \item The library can be used directly in a quantum cryptographic protocol to \textbf{implement the privacy amplification step}.
    
    \item The library can be used to test and \textbf{validate high-performance implementations}, whose source code may be unavailable or difficult to audit. We use \texttt{randextract} to test and validate three randomness extractors that have been used in recent QKD and QRNG implementations. With these real world examples, we highlight the importance of the classical post-processing phase for security and correctness.

    \item We provide a function to generate \textbf{test vectors} for privacy amplification implementations in order to facilitate external audits. We use the same format as in the NIST Cryptographic Algorithm Validation Program. These test vectors can be used by the community in a validation process included into future standardization efforts of quantum cryptographic devices.
\end{enumerate}

\subsection{Outline}\label{sec:outline}

The paper is organized as follows: Sec.~\ref{sec:theoretical_background} introduces the background of quantum key distribution, quantum random number generation and privacy amplification. In particular, Sec.~\ref{sec:randomness-extractors} introduces the theory of two families of quantum-proof extractors: (modified) Toeplitz hashing and Trevisan's construction. Sec.~\ref{sec:randextract}, \ref{sec:validating}, and \ref{sec:test-vectors} present our main contributions. In Sec.~\ref{sec:randextract}, we describe our Python package, \texttt{randextract}, in detail. Sec.~\ref{sec:validating} demonstrates how the library can be used to test and validate third-party implementations. Notably, using our library, we identify and correct a bug in a modified Toeplitz hashing implementation used in a high-speed QKD experiment. We also uncover discrepancies between the mathematical specification and the actual implementation of a high-performance Trevisan's construction implementation used in QKD and QRNG experiments.

Sec.~\ref{sec:test-vectors} introduces a set of test vectors for Toeplitz hashing, proposed as a step towards future standardization efforts. Finally, Sec.~\ref{sec:conclusion} presents our conclusions and outlines directions for future work.

%%
%% 2. THEORETICAL BACKGROUND
%%
\section{\uppercase{Theoretical background}}\label{sec:theoretical_background}

\subsection{Notation \& Definitions}
\label{sec:notation}

The capital letters $X$, $Y$ and $Z$ denote \emph{classical random variables}, which take values $x$, $y$ and $z$. Calligraphic fonts $\mathcal{X}$, $\mathcal{Y}$ and $\mathcal{Z}$ denote the \emph{alphabet} of the corresponding random variables. The \emph{probability} that the random variable $Z$ takes the value $z$ is $P_Z(Z=z)$, sometimes abbreviated to $P_Z(z)$ or $P(z)$ when it is clear from the context which random variable it refers to. For two random variables $X$ and $Y$, the \emph{joint probability} is defined as $P_{XY}(x,y)=P(X=x \land Y=y)$, and the \emph{conditional probability} of $X$ given $Y$ as $P_{X|Y}(x,y)=P_{X|Y=y}(x)=P_{XY}(x,y)/P_Y(y)$.

The \emph{guessing probability}, defined as the probability of correctly guessing the value of a random variable $X$, is
\begin{equation}\nonumber
p_\text{guess}(X) = \max_{x\in\mathcal{X}}P_X(X=x)\,,
\end{equation}
which motivates the definition of the \emph{min-entropy}
\begin{equation}
\nonumber
H_\text{min}(X)=-\log_2 p_\text{guess}(X)\,.
\end{equation}

The \emph{uniform} distribution over a random variable $X$ is denoted by $U_X$, i.e., each $x\in\mathcal{X}$ is equally likely, and the guessing probability is $1/|\mathcal{X}|$.

The \emph{statistical distance} between two probability distributions $P_X$ and $P_Y$, defined over the same alphabet $\mathcal{X}$, is given by
\begin{equation}\nonumber
d(X, Y) = \frac{1}{2} \sum_{x\in\mathcal{X}}\left|P_X(x=x) - P_Y(y=x)\right|\,.
\end{equation}
Operationally, the statistical distance represents the maximum advantage one can gain in correctly identifying whether a single sample was drawn from $P_X$ or $P_Y$. A probability distribution of a random variable $X$ is said to be \emph{$\epsilon-$close to uniform} if its distance with respect to the uniform distribution is bounded by $\epsilon$, i.e.,
\begin{equation}\label{eq:stat_dist}
d(X, U) = \frac{1}{2}\sum_{x\in\mathcal{X}}\left|P(x)-\frac{1}{|\mathcal{X}|}\right|
\leq\epsilon\,.
\end{equation}

Greek letters, such as $\rho$ and $\sigma$, are used to denote quantum states. Given a Hilbert space $\mathcal{H}$, a \emph{quantum state} $\rho$ can be represented by a positive semi-definite Hermitian operator of unit trace acting on $\mathcal{H}$. Any quantum state can be expressed in terms of an orthonormal basis $\{\ket{i}\}_{i}$ of $\mathcal{H}$
\begin{equation}\nonumber
\rho = \sum_{i,j}\rho_{ij}\ket{i}\bra{j}\,,
\end{equation}
where $\rho_{ij}$ are complex numbers, $\rho_{ij}=\rho_{ji}^*$, and $\Tr[\rho]=\sum_i\rho_{ii}=1$.

Letters from the beginning of the alphabet, such as $A$, $B$ or $E$ are used to denote the subsystems that form a \emph{composite quantum system}. For example, $\rho_{ABE}$ denotes a quantum state acting on the Hilbert space $\mathcal{H}=\mathcal{H}_A\otimes\mathcal{H}_B\otimes\mathcal{H}_E$. 

A (classical) random variable $X$ can be written as a \emph{diagonal} quantum state
\begin{equation}\nonumber
\rho_X=\sum_{x\in\mathcal{X}}P(x)\ketbra{x}\,,
\end{equation}
and the uniform distribution can be represented as the \emph{maximally mixed state}
\begin{equation}\nonumber
\sigma_U=\frac{1}{\dim(\mathcal{H})}\sum_{i=1}^{\dim(\mathcal{H})}\ketbra{i}\,.
\end{equation}

Of special interest in quantum cryptography are the so-called \emph{classical-quantum (cq) states}, which allow to represent a quantum system correlated with a classical random variable. Given a probability distribution $P_X$ and a set of quantum states $\{\rho_E^x\}_x$, a cq-state is defined as
\begin{equation}\label{eq:cq-state}
\rho_{XE}=\sum_{x\in\mathcal{X}}\left(P_X(x)\ketbra{x}\otimes\rho_{E}^x\right)\,.
\end{equation}

This formalism is particularly relevant in security analyses, where the classical variable models the honest party's output (e.g., a raw key), and the quantum system $E$ represents the adversary's \emph{side information}.

\emph{Quantum measurements} are represented by a set of positive operators $\{E_i\}_i$ satisfying $\sum_i E_i=\mathds{1}$. These sets are called \emph{Positive Operator-Valued Measure} (POVM).
Given a state $\rho$ and a measurement $\{E_i\}_i$, the probability of measuring the value $i$ is given by $\Tr[E_i\rho]$.

The \emph{trace distance} generalizes the notion of statistical distance to the quantum setting. For two quantum states $\rho$ and $\sigma$, it is defined as
\begin{equation}\nonumber
d(\rho,\sigma)=\frac{1}{2}\left\|\rho-\sigma\right\|_\text{tr}=\frac{1}{2}\Tr\sqrt{(\rho-\sigma)^\dag(\rho-\sigma)}\,.
\end{equation}

Analogously to the statistical distance, the trace distance provides an operational interpretation as a measure of distinguishability between quantum states. If a system is prepared in either $\rho$ or $\sigma$ with equal probability, the maximum probability with which the state can be correctly identified; e.g., by doing a (single) quantum measurement or feeding the state into another quantum protocol, is $\frac{1}{2} + \frac{1}{2}d(\rho,\sigma)$.

The trace distance of a classical random variable $X$ from uniform from the point of view of an adversary with quantum side information $E$ is given by
\begin{equation}\label{eq:trace_dist_uniform}
d(\rho_{XE},\sigma_U\otimes\rho_E)=\frac{1}{2}\left\|\rho_{XE}-\sigma_U\otimes\rho_E\right\|_\text{tr}\,.
\end{equation}

Finally, both the guessing probability and the min-entropy can be generalized to the setting of cq-states. The guessing probability of a random variable $X$ conditioned on a quantum system $E$ is
\begin{equation}\nonumber
p_\text{guess}(X|E)=\max_{\{E_x\}}\sum_{x\in\mathcal{X}}P_X(x)\Tr[\rho_E^xE_x]\,,
\end{equation}
where the maximization is over all the POVMs, and the conditional min-entropy
\begin{equation}\nonumber
H_\text{min}(X|E)=-\log_2 p_\text{guess}(X|E)\,.
\end{equation}

\subsection{Quantum Cryptography}\label{sec:quantum_crypto}

\subsubsection{Quantum Key Distribution}\label{sec:qkd}

An \emph{ideal} key distribution protocol is one that produces a bit string that is identical for the honest parties and that looks uniform from the adversary's point of view. \emph{Real} protocols cannot usually achieve this perfectly and, instead, produce an output that is $\epsilon$-close to the ideal one, i.e., the trace distance between the output of the real and the ideal protocol, as in Eq.~\eqref{eq:trace_dist_uniform}, is bounded by $\epsilon$. The trace distance guarantees that the real protocol is indistinguishable from the ideal protocol, except with probability $\epsilon$. This notion of security as the distinguishability between an ideal and a real protocol guarantees that the protocol is \emph{composable}~\cite{pw,bpw,canetti,Maurer02}, i.e., it remains secure even if it is used with other protocols to form a larger cryptographic system.

Key distribution is traditionally implemented using public-key cryptography, e.g., RSA~\cite{rsa}, Diffie-Hellman~\cite{dh} or ML-KEM~\cite{kyber,mlkem}. These protocols are secure if the adversary is limited in computing power, and given some assumptions on the computational hardness of certain mathematical problems.

Quantum Key Distribution (QKD) is a quantum cryptographic protocol that enables two honest parties to establish a shared secret key over an insecure quantum channel\footnotemark.  QKD is fundamentally different from public-key cryptography because the established keys are information-theoretically secure~\cite{Maurer99}, i.e., their security does not depend on a limitation in computing power of the adversary.

\footnotetext{In addition to the public quantum channel, the honest parties need to be able to communicate classically over an authentic channel and have access to local randomness. Their labs need to be secured and isolated, and depending on the protocol, there might be additional conditions on the quantum devices.}

Any quantum key distribution protocol consists of two distinct phases:

\begin{enumerate}
    \item a quantum phase, in which quantum states are prepared, transmitted over a public quantum channel, and measured;
    \item and a classical phase, in which the classical data that corresponds to the preparation settings and measurement results is post-processed to derive the final shared secret keys.
\end{enumerate}

The classical post-processing involves several subprotocols or steps, which might differ slightly from one particular protocol to another. A common way to implement this phase in a QKD protocol is the following:
\begin{enumerate}
    \item The starting point after the quantum phase are two \emph{raw keys}, bit strings held by the honest parties that are neither equal nor completely secret.
    \item A sample from the raw keys is compared and then discarded to perform the \emph{testing} step, also known as \emph{parameter estimation}. The honest parties reveal some of the bits over the classical channel to estimate the losses and number of errors, which allows to bound the amount of information an adversary might have obtained during the quantum phase\footnotemark. Testing aborts the protocol if this estimation is beyond a certain threshold, and continues otherwise. A real implementation correctly recognizes when an adversary has too much information except with probability $\epsilon_\text{pe}$. 
    \item In case the honest parties decide to continue the protocol, the next step is called \emph{information reconciliation} and corrects the errors of the (remaining) raw keys to obtain the \emph{corrected keys}. This step ensure the correctness of the protocol, and it is often implemented using error-correcting codes (see e.g.,~\cite{PetersonWeldon1972}). After this step, the two honest parties hold identical keys, which are partially secret, except with probability $\epsilon_\text{ir}$.
    \item Lastly, the corrected keys are transformed into shorter but completely secret bit strings in the \emph{privacy amplification} step. These are the final \emph{secret keys}. At the end of this step, the honest parties still hold equal keys but these are now also secret from the adversary except with probability $\epsilon_\text{pa}$.
\end{enumerate}

\footnotetext{In reality, errors and losses might be due to the noisy quantum channel. In cryptography, it is common to assume the worst-case scenario. In this case, any error is attributed to an adversary eavesdropping and manipulating the transmitted quantum states.}

A \emph{security proof} of a QKD protocol~\cite{rennerphd} is a theoretical statement, with well-defined conditions and a security claim in terms of a \emph{security parameter} $\epsilon$ bounding the probability that the output of the protocol differs from an ideal key distribution protocol, i.e., guarantees that the honest parties end up with an equal, perfectly secure key except with probability $\epsilon$. If the subprotocols are defined in a composable way, then the security parameter can be bounded in terms of the parameters of these subprotocols using the union bound~\cite{Boole1847}, i.e., $\epsilon\leq\epsilon_\text{pe}+\epsilon_\text{ir}+\epsilon_\text{pa}$. This security parameter $\epsilon$ can ideally be made arbitrarily small at the cost of a lower key rate.

\subsubsection{Quantum Random Number Generator}\label{sec:qrng}

A Quantum Random Number Generator (QRNG) is a type of hardware random number generator that uses quantum systems to generate randomness. Most hardware random number generators use classical physical processes, and since these processes are deterministic, these protocols need to assume that the initial state of the system is only partially known by an adversary. QRNGs, on the other hand, take advantage of the inherent unpredictability of quantum phenomena. Even if a quantum state is perfectly described, the outcome of a measurement is, in general, not deterministic~\cite{Born1955}.

QRNG protocols are similar to QKD protocols but with just one party involved. In the quantum phase, the same party prepares and measures quantum states, and the classical post-processing does not contain an information reconciliation step. The security notion for QRNG protocols is the same as in QKD, i.e., close-to-uniform randomness even from the point of view of an adversary who might hold a quantum state initially correlated with the state of the QRNG system. The privacy amplification step can be used in exactly the same way as in a QKD protocol. It removes bias arising from the quantum phase and side-information held by an adversary.

\subsection{Privacy Amplification with Seeded Randomness Extractors}\label{sec:PA}

As introduced in Sec.~\ref{sec:qkd}, Privacy Amplification (PA) is a classical post-processing procedure whose goal is to \emph{compress} partially secret (classical) bit strings into bit strings that are uniform and independent of any adversary's (quantum) knowledge. In the context of QKD, the adversary's knowledge of the corrected keys can originate from eavesdropping or tampering with the quantum states transmitted over the public quantum channel, as well as from observing the authenticated classical communication, such as the error correction syndrome exchanged during the information reconciliation step.

The correct composable definition of ideal keys or ideal randomness after privacy amplification is an upper bound on Eq.~\eqref{eq:trace_dist_uniform}, i.e.,
\begin{equation}\nonumber
d(\rho_{XE},\sigma_U\otimes\rho_E)\leq\epsilon_\text{pa}\,,
\end{equation}
which guarantees that the final keys from QKD or the output from a QRNG look uniform from an adversary's point of view, except with a small probability $\epsilon_\text{pa}$.

\begin{figure}
    \centering
    \includegraphics[width=\linewidth]{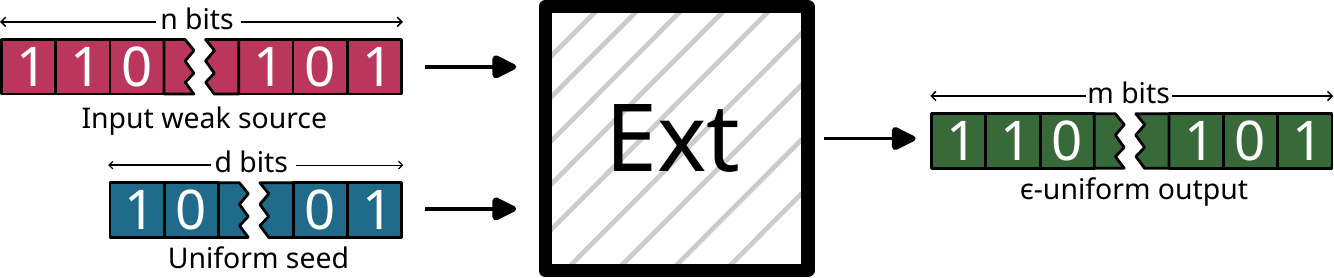}
    \caption{Diagram of a seeded randomness extractor that takes an $n$-bit string and a uniform $d$-bit uniform string as inputs, and outputs an $\epsilon$-close to uniform $m$-bit string.}
    \label{fig:seeded_extractor}
\end{figure}

\paragraph{Seeded Randomness Extractors}
Privacy amplification can be implemented using seeded randomness extractors (see e.g.,~\cite{SHALTIEL2004}). A \emph{seeded randomness extractor} is a function that takes as input
\begin{enumerate}
    \item a weak random source $X$, i.e., a bit string of length $n$, and
    \item a uniform seed $Y$, i.e., a bit string of length $d$,
\end{enumerate}
and outputs a bit string of length $m$ (see Figure \ref{fig:seeded_extractor}), i.e.,
\begin{equation}\label{eq:seeded_ext}
\text{Ext}\::\: \{0, 1\}^n \times \{0, 1\}^d \rightarrow \{0, 1\}^m
\end{equation}

A seeded randomness extractor can be used to transform a bit string with high min-entropy into an almost-uniform bit string. In order to correctly implement the privacy amplification step using these extractors we require two additional conditions:
\begin{enumerate}
    \item The output of the extractor is independent of the seed. This is because the random seed is revealed to the adversary during the execution of the protocol.
    \item The output is close to uniform for an adversary with quantum side information.
\end{enumerate}

Extractors satisfying the first condition are called \emph{strong}, and those satisfying the second condition \emph{quantum-proof}\footnotemark.

\footnotetext{If the output of the extractor is close to uniform for an adversary with classical side information, the extractor is said to be \emph{classical-proof}. All quantum-proof extractors are classical-proof, but the converse is not true.}

A quantum-proof ($k,\epsilon)$-strong seeded randomness extractor~\cite{KonigRenner2011} is a function defined as in Eq.~\eqref{eq:seeded_ext} which, for a uniform independent seed $Y$ and input $X$ with conditional min-entropy $H_\text{min}(X|E)\geq k$, outputs a string $\text{Ext}(X, Y)$ satisfying
\begin{equation}\label{eq:dist_quantum_proof_ext}
d(\rho_{\text{Ext}(X, Y)YE}, \sigma_U\otimes\rho_Y\otimes\rho_E) \leq \epsilon_\text{pa}\,,
\end{equation}
i.e., the output is $\epsilon_\text{pa}$-close to the ideal output which is uniform, independent of the seed and uncorrelated to the quantum side information.

In the context of QKD and QRNG protocols, only quantum-proof strong randomness extractors can to be used to realize the privacy amplification step. Therefore, these are the extractors implemented in \texttt{randextract}.

\subsection{Construction of Randomness Extractors}\label{sec:randomness-extractors}
Families of two-universal hash functions are quantum-proof strong extractors. Alternative constructions are possible such as the Trevisan's construction.

A family of functions $\mathcal{F}=\{f_y\}_y$, such that $f_y\::\:\mathcal{X}\rightarrow\mathcal{Z}$, is \emph{two-universal} if
\begin{equation}\nonumber
\Pr_{f_y\in\mathcal{F}}[f_y(x)=f_y(x')]\leq\frac{1}{|\mathcal{Z}|}\,,
\end{equation}
for any distinct $x,x'\in\mathcal{X}$, and $f_y$ picked uniformly at random~\cite{CarterWegman1979}. This means that the number of collisions, when the function is picked uniformly at random, is bounded by the size of the output alphabet. There exist two-universal functions from $\{0, 1\}^n$ to $\{0, 1\}^m$ for any positive integers $0\leq m \leq n$~\cite{WegmanCarter1981}.

A seeded randomness extractor, as defined in Eq.~\eqref{eq:seeded_ext}, is realized from a family of two-universal hash functions $\mathcal{F}$ by using the seed $y\in\mathcal{Y}$ to choose one particular function $f_y\in\mathcal{F}$ and hashing the input $x\in\mathcal{X}$ with this chosen function, i.e.,
\begin{equation}\label{eq:ext_from_universal}
\text{Ext}(x, y):= f_y(x)\,.
\end{equation}

The quantum leftover hash lemma~\cite{rennerkoenig,rennerphd,qlhl} states that a family of two-universal functions, as defined in Eq.~\eqref{eq:ext_from_universal}, is a quantum-proof $(k, \epsilon_\text{pa})$-strong extractor for output length $m$ less or equal than
\begin{equation}\nonumber
\left\lfloor k + 2 - 2\log \frac{1}{\epsilon}\right\rfloor\,.
\end{equation}

\subsubsection{Toeplitz Hashing}

\begin{figure}
    \centering
    \includegraphics[width=0.8\linewidth]{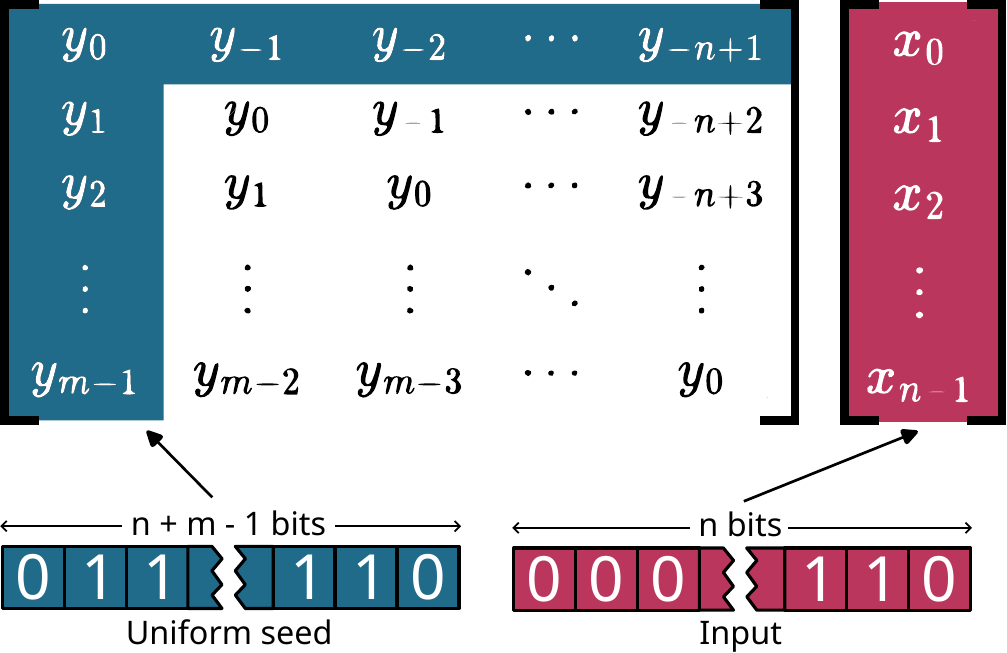}
    \caption{Toeplitz hashing is defined as the matrix-vector multiplication of a Toeplitz matrix generated from a uniform seed $y$ of length $n+m-1$ and input $x$ of length $n$ from a weak random source.}
    \label{fig:toeplitz_hashing}
\end{figure}

The set of all binary $m\times n$ Toeplitz matrices defines a family of two-universal hash functions, where each matrix corresponds to a specific function in the family. A matrix $T$ is said to be a Toeplitz matrix if its elements satisfy
\begin{equation}\label{eq:Toeplitz_matrix}
T_{i,j}=T_{i-1,j-1}\,.
\end{equation}
Toeplitz matrices are, therefore, fully characterized by their first row and column. 

% Hashing is performed via matrix-vector multiplication: the input $x\in\mathds{F}_2^n$ is interpreted as an $n\times1$ column vector, and the selected Toeplitz matrix is applied from the left to produce an output vector in $\mathds{F}_2^m$.

One common method to select a specific matrix from this family using the seed $y$ is to define the matrix entries based on $y$ as follows:
\begin{equation}\label{eq:Toeplitz_matrix_seed_construction}
T_{i,j}(y):= y_{i-j}\,.
\end{equation}

\paragraph{Standard Toeplitz Hashing} Using Eq.~\eqref{eq:Toeplitz_matrix_seed_construction} to construct a matrix from the seed, the standard Toeplitz hashing (see e.g.,~\cite{Krawczyk1995}) is defined as
\begin{equation}\label{eq:ext_toeplitz}
\text{Ext}_\text{\tiny Toeplitz}(x, y):=T(y)\cdot x\,.
\end{equation}
Figure \ref{fig:toeplitz_hashing} shows this construction graphically.

\paragraph{Efficient implementation} General matrix-vector multiplication has a computational complexity $\mathcal{O}(n^2)$. However, the matrix-vector multiplication of Toeplitz hashing can be implemented more efficiently with a computational complexity $\mathcal{O}(n\log n)$ using the Fast Fourier Transform (FFT). First, the $m\times n$ Toeplitz matrix is transformed into a square $(m+n-1)\times(m+n-1)$ circulant matrix by adding $n-1$ additional rows and $m-1$ columns. Then, this matrix can be diagonalized using the Fourier matrix $F_q$, whose matrix elements are determined by $F_{j,k}=\frac{1}{\sqrt{q}}e^{2\pi ijk/q}$ with $q=m+n-1$. The circulant matrix can, therefore, be written in terms of the seed $y$ as
\begin{equation}\nonumber
\hat{T}_q(y)=F_q^{-1}\text{diag}(F_q y)F_q\,.
\end{equation}

And the complete Toeplitz extractor as
\begin{equation}\nonumber
\text{Ext}_\text{\tiny Toeplitz}(x,y)=\text{FFT}^{-1}\Big(\text{FFT}(y) \odot \text{FFT}(\hat{x})\Big)\Big|_{0\dots m-1}\,,
\end{equation}
where $\hat{x}$ is the input $x$ padded with $m-1$ zeros, $\odot$ denotes element-wise multiplication of the two vectors, and $|_{0\dots m-1}$ means that the output of the inverse fast Fourier transform is truncated to the first $m$ bits (see App.~C from~\cite{Hayashi2016}).

\paragraph{Modified Toeplitz Hashing} 
Standard Toeplitz hashing requires a seed of length $m + n - 1$ bits. However, it is possible to reduce the seed length to $n - 1$ bits by using a different family of two-universal hash functions~\cite{Hayashi2016}. For the same input and output lengths, the so-called \emph{modified Toeplitz hashing} uses as its hashing matrix the concatenation of a smaller Toeplitz matrix with the $m \times m$ identity matrix, i.e.,
\begin{equation}\label{eq:ext_mod_toeplitz}
\text{Ext}_\text{\tiny Mod.~Toeplitz}(x, y) := (T'(y) \,\|\, \mathds{1}_m)x\,,
\end{equation}
where $T'(y)$ is a Toeplitz matrix of dimension $m \times (n - m)$. The matrix-vector multiplication remains efficiently computable using the same FFT-based approach described above.

\subsubsection{Trevisan's Construction}

Two-universal hashing is not the only way to define strong quantum-proof seeded extractors. Trevisan~\cite{trevisan}  developed a method to construct arbitrary extractors from one-bit extractors and weak designs. Later, it was proven that this construction generates a quantum-proof strong extractor~\cite{deportmannvidickrenner} from a strong one-bit extractor.

\paragraph{One-Bit Extractors} One-bit extractors are simply extractors defined as in Eq.~\eqref{eq:seeded_ext} with $m=1$. Any one-bit $(k,\epsilon)$-strong extractor is a quantum-proof $(k-\log\epsilon, 3\sqrt{\epsilon})$-strong extractor. If the one-bit extractor is already classical-proof, then it can be shown that it is also a quantum-proof $(k,(1+\sqrt{2})\sqrt{\epsilon})$-strong extractor~\cite{koenigterhal}.\\

A combinatorial design~\cite{Nisan1994} is a family of subsets $W=[S_0,S_1,\dots,S_{m-1}]$, with $S_i\subseteq[d]$, typically constructed so that the intersections between the subsets satisfy certain constraints. These designs are widely used in theoretical computer science due to their ability to balance overlap and independence.

\paragraph{Weak Design} A weak $(m,t,r,d)$-design is a combinatorial design, i.e., a family of sets $W=[S_0, S_1,\dots,S_{m-1}]\in[d]$, where all sets are of size $t$, satisfying
\begin{equation}\label{eq:weak-design-bound}
\sum_{j=0}^{i-1}2^{|S_i\cap S_j|}\leq rm
\end{equation}
for all $i$~\cite{Raz2002}.

\begin{figure}
    \centering
    \includegraphics[width=\linewidth]{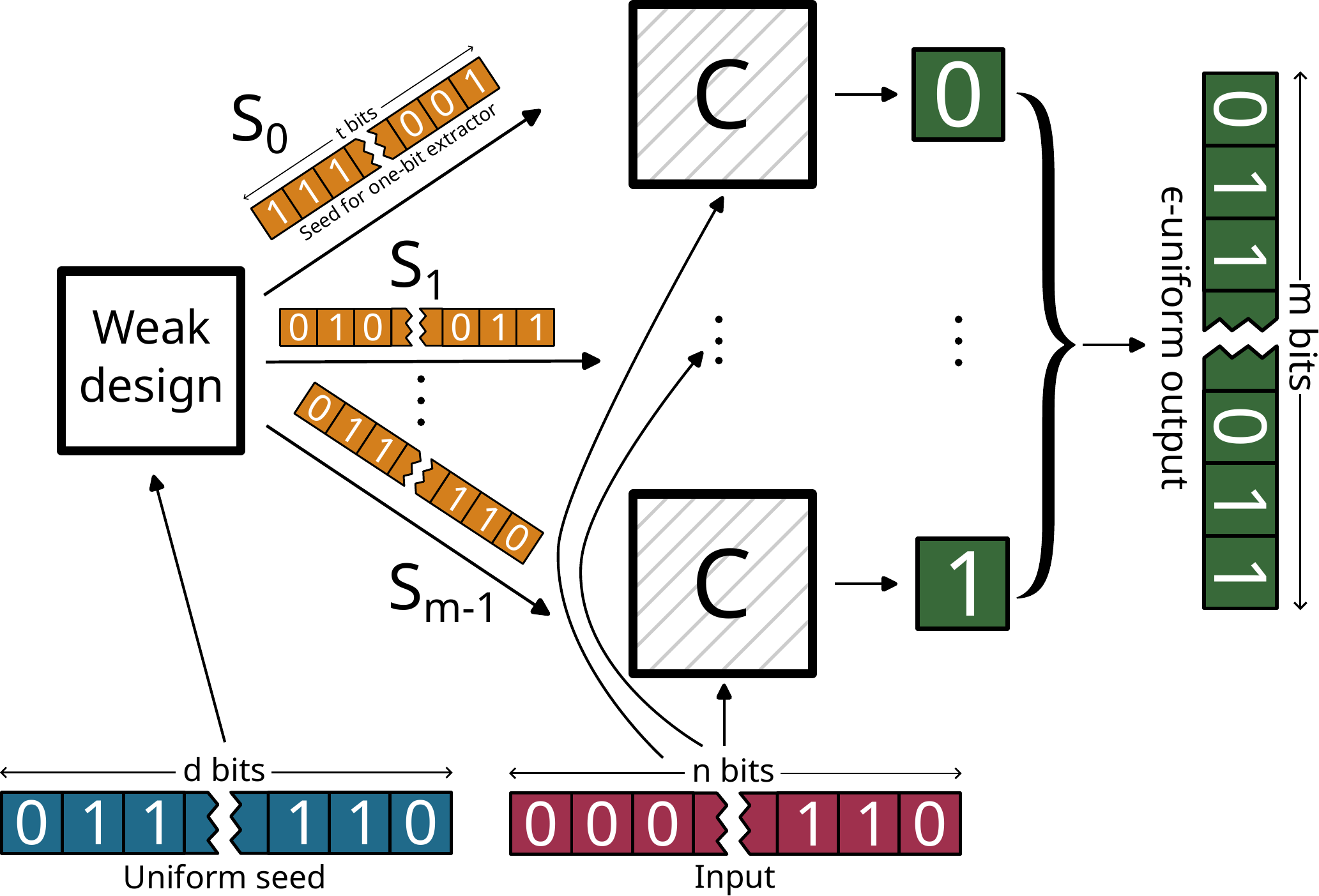}
    \caption{Trevisan's extractor output \eqref{eq:trevisan_construction} is the concatenation of the $m$ bits obtained from calling the one-bit extractor $m$ times on the same input but with a different seed each time. The $m$ seeds $y_{S_0},\dots,y_{S_{m-1}}$ are constructed from the input seed $y$ using the sets from the weak design as indices.}
    \label{fig:trevisan_construction}
\end{figure}

\paragraph{Trevisan's extractor}
Given a quantum-proof strong one-bit extractor $\text{Ext}_1(x, y)$ and a weak design $W=[S_0,\dots,S_{m-1}]$, Trevisan's extractor is realized as
\begin{equation}\label{eq:trevisan_construction}
\text{Ext}_\text{\tiny Trevisan}(x, y):=\text{Ext}_1(x, y_{S_0})\dots \text{Ext}_1(x, y_{S_{m-1}})\,.
\end{equation}
Figure \ref{fig:trevisan_construction} shows this construction graphically.

\paragraph{Computational complexity}
The computational complexity of Trevisan's construction depends on the specific choice of the weak design and the one-bit extractor. In practice, Trevisan's extractor is often significantly slower than Toeplitz hashing. However, it offers an important advantage: certain constructions require only a seed of polylogarithmic length in the input size. For instance, the construction using a weak design over a finite field and a polynomial-time one-bit extractor requires a seed of length $\mathcal{O}(\log^2 n)$~\cite{mauerer2012modularframeworkrandomnessextraction}. In contrast, Toeplitz and modified Toeplitz hashing require seeds of length $\mathcal{O}(n)$. Trevisan's extractor may therefore be preferable in scenarios where good randomness is a scarce resource.

%%
%% 3. RANDEXTRACT
%%
\section{\uppercase{randextract}}\label{sec:randextract}

\texttt{randextract} is our open-source Python package implementing some of the most used privacy amplification algorithms. It is available at the Python Package Index (PyPI)~\cite{randextract-pypi}. We recommend installing it in a virtual Python environment, for example\footnotemark, by running the following commands

\begin{lstlisting}[]
python -m venv --upgrade-deps env-randextract
source env-randextract/bin/activate
pip install randextract
\end{lstlisting}

\footnotetext{In Windows, the activation of the virtual environment should be done running \lstinline|env-randextract\Scripts\Activate.ps1| instead.}
Other ways of installing the package are described in the latest available online documentation\footnotemark.
\footnotetext{\url{https://randextract.crypto-lab.ch}}

\subsection{Overview \& Goals}

The goal of \texttt{randextract} is to provide a readable and easy-to-audit library implementing most of the PA algorithms used in quantum cryptographic protocols. The source code stays close to the mathematical formulation to avoid deviations and facilitate the audit of its correctness.  Only when it can be done without impacting this primary goal, secondary goals such as performance are taken into consideration. The package also provides classes and helping functions to aid validating third-party implementations. The package is intended to be self-contained, with comprehensive documentation that introduces the theory of randomness extractors, explains their relevance to quantum cryptographic protocols, includes several toy examples suitable for manual computation, and provides real-world validation cases.

\subsection{Structure \& Design}

The source code repository follows a standard Python structure.

\begin{itemize}
    \item \texttt{src/randextract}: source code of the library.
    \item \texttt{tests}: unit and integration tests.
    \item \texttt{docs/source}: source code of the online documentation\footnotemark[5].
    \item \texttt{examples}: scripts validating real world privacy amplification implementations.
    \item \texttt{resources}: additional resources such as plots, datasets used in testing and the scripts to generate them, Jupyter notebooks, test vectors, etc.
\end{itemize}

The package is modular by design allowing to be easily extended in the future with additional extractors. A common API is enforced using abstract classes. For example, the abstract class \lstinline|RandomnessExtractor| defined in the module \lstinline|randomness_extractor.py| requires that the implementation classes, such as the \lstinline|ToeplitzHashing| class, implement the properties \lstinline|input_length|, \lstinline|seed_length| and \lstinline|output_length|, and the method \lstinline|extract()|.

\subsection{Dependencies}
\texttt{randextract} uses the Python Standard Library~\cite{python-sl} and the well-known numerical libraries NumPy~\cite{Harris2020} and Scipy~\cite{Virtanen2020}. In addition, the Galois package~\cite{Hostetter2020} is used to compactly define arrays and polynomials over finite fields.

\subsection{Usage}

\texttt{randextract} can be used directly to implement the PA step of quantum cryptographic protocols. It can also be used to validate other implementations. In both cases the workflow is the same:

\begin{enumerate}
    \item Choose a particular class of seeded randomness extractors.
    \item Instantiate one particular extractor with the desired input and output lengths, and the required parameters. When using \texttt{randextract} to implement the PA step, the optimal parameters can be calculated with the library. When validating other implementations, these have to match the implementation being tested.
    \item Extract the randomness from the weak source, or validate the implementation.
\end{enumerate}

\paragraph{Implementing PA}
The choice of the extractor class in Step 1 depends on the requirements of the quantum cryptographic protocol. In Step 2, the optimal input and output lengths can be computed using \texttt{randextract}, based on the model of the weak randomness source and the desired security parameter~$\epsilon_\text{pa}$ from Eq.~\eqref{eq:dist_quantum_proof_ext}. This is handled by the \lstinline|calculate_length()| helper method available in any \lstinline|RandomnessExtractor| implementation. Finally, the output is computed in Step 4 by calling the \lstinline|extract()| method on the instantiated extractor.

\paragraph{Validating other implementations}
To validate an external implementation, Steps 1 and 2 must follow the same extractor class and parameter choices. In Step 3, rather than manually extracting outputs and comparing them across different seeds and inputs, the custom extractor can be passed directly to the \lstinline|Validator| class. External implementations are registered using the \lstinline|add_implementation()| method, and validation is performed via the \lstinline|validate()| method. Any failing test cases are recorded, and detailed insights into discrepancies can be obtained using the \lstinline|analyze_failed_test()| method.

\subsection{Code Examples}

\subsubsection{Toeplitz Hashing}

\paragraph{Implementing PA with Toeplitz hashing}
The following code snippet illustrates how to create an extractor based on standard Toeplitz hashing, as defined in Eq.~\eqref{eq:ext_toeplitz}. First, the relevant classes are imported:
\begin{lstlisting}[firstnumber=1]
from galois import GF2

import randextract
from randextract import (
    RandomnessExtractor,
    ToeplitzHashing
)
\end{lstlisting}

Next, the optimal output length is computed for inputs of 8~Mib originating from a weak random source with initial min-entropy $\frac{1}{n}H_\text{min}(X|E)\geq\frac{1}{2}$, and for a security parameter of $\epsilon_\text{pa}=10^{-6}$.
\begin{lstlisting}[firstnumber=8]
Mib = 2**20

out_len = ToeplitzHashing.calculate_length(
    extractor_type="quantum",
    input_length=8*MiB,
    relative_source_entropy=0.5,
    error_bound=1e-6,
)
\end{lstlisting}

A Toeplitz hashing extractor is then instantiated for the corresponding input and output lengths:
\begin{lstlisting}[firstnumber=16]
ext = RandomnessExtractor.create(
    extractor_type="toeplitz",
    input_length=8*MiB,
    output_length=out_len
)
\end{lstlisting}

Finally, the hashing is applied to (pseudo)random input and seed bit strings:
\begin{lstlisting}[firstnumber=21]
ext_int = GF2.Random(MiB)
ext_seed = GF2.Random(ext.seed_length)
ext_out = ext.extract(ext_int, ext_seed)
\end{lstlisting}

\paragraph{Validating a third-party implementation of modified Toeplitz hashing}
The following code snippet demonstrates how to validate a third-party implementation. First, two additional classes are imported:
\begin{lstlisting}[firstnumber=1]
from randextract import (
    ModifiedToeplitzHashing,
    Validator
)
\end{lstlisting}

The reference extractor is instantiated using the \lstinline|ModifiedToeplitzHashing| class, which implements Eq.~\eqref{eq:ext_mod_toeplitz}, with the same input and output lengths as the implementation under test. In this case, the extractor takes inputs of 1~Mib and compresses them by 50\%. The resulting object is functionally equivalent to one created via \lstinline|RandomnessExtractor.create()|, as shown in the previous example.
\begin{lstlisting}[firstnumber=5]
ref_ext = ModifiedToeplitzHashing(
    input_length=2**20,
    output_length=2**19
)
\end{lstlisting}

An auxiliary function is defined to convert binary arrays into string representations:
\begin{lstlisting}[firstnumber=9]
def gf2_to_str(gf2_arr):
    arr = np.array(gf2_arr)
    arr_str = (arr + ord("0")).tobytes()
    return arr_str.decode()
\end{lstlisting}

A \lstinline|Validator| instance is then initialized with the reference extractor, and the third-party implementation is registered. The validator is configured to interact with the implementation via standard input/output, using the previously defined conversion function as a parser:
\begin{lstlisting}[firstnumber=13]
val = Validator(ext)
val.add_implementation(
    label="Rust-stdio-fft",
    input_method="stdio",
    command="./modified_toeplitz $SEED$ $INPUT$",
    format_dict={
        "$SEED$": gf2_to_str,
        "$INPUT$": gf2_to_str
    },
)
\end{lstlisting}

Validation is performed using $10^4$ randomly generated input samples:
\begin{lstlisting}[firstnumber=23]
val.validate(
    mode="random",
    sample_size=10**4
)
\end{lstlisting}

\subsubsection{Trevisan's extractor}

To instantiate a randomness extractor based on Trevisan's construction, additional parameters must be specified. In addition to the input and output lengths, common to all seeded extractors, one must select concrete implementations for the one-bit extractor and the weak design. These implementations may impose constraints on the parameters. For instance, the weak design provided by the class \lstinline|FiniteFieldPolynomialDesign| requires the size of the subsets to be a prime number (or a prime power). Such constraints are automatically handled by the \lstinline|calculate_length()| method.

The following example shows how to instantiate a Trevisan extractor configured to take 1~Mib of input and output 1~Kib, using a finite field weak design and a polynomial-based one-bit extraction. No additional imports are required:
\begin{lstlisting}[firstnumber=1]
ext = RandomnessExtractor.create(
   extractor_type="trevisan",
   weak_design_type="finite_field",
   one_bit_extractor_type="polynomial",
   one_bit_extractor_seed_length=1024,
   input_length=2**20,
   output_length=2**10,
)
\end{lstlisting}

Once the extractor object has been instantiated, the extraction process is performed identically across all implementations:
\begin{lstlisting}[firstnumber=9]
ext_input = GF2.Random(ext.input_length)
ext_seed  = GF2.Random(ext.seed_length)
ext_out = ext.extract(ext_input, ext_seed)
\end{lstlisting}

Additional examples and the complete API documentation for \texttt{randextract} are available online\footnote{\url{https://randextract.crypto-lab.ch/api.html}}.

%%
%% 4. VALIDATING
%%
\section{\uppercase{Validating privacy amplification implementations}}
\label{sec:validating}

\texttt{randextract} is well-suited for small proof-of-concept implementations or protocols where privacy amplification can be performed offline. In contrast, production environments requiring real-time privacy amplification typically demand performance-optimized and highly efficient implementations. In the last years, due to huge improvements on the quantum-physical part of quantum cryptography protocols~\cite{fadriphd}, the classical post-processing has become a bottleneck~\cite{10mbitQKD}. The secret key rate of QKD protocols and the throughput of QRNGs is limited by how fast the classical post-processing can be done. This has motivated the development of high-performance implementations of both information reconciliation and PA algorithms, using hardware accelerators such as GPUs~\cite{Bosshard2021} or FPGAs~\cite{ToeplitzFPGA}. These implementations are harder to read and audit, and deviations from the mathematical definitions are harder to spot.

In this section we show how we used \texttt{randextract} to test, validate and fix such high-performance PA implementations.

\subsection{GPU modified Toeplitz hashing}

Modern GPUs can accelerate a wide range of computational tasks, including the calculation of the FFT. A CUDA and Vulkan-based implementation of the modified Toeplitz hashing algorithm~\cite{Bosshard2021} was employed in a QKD experiment~\cite{nphotonics} involving ultrafast single-photon detectors, where constraints on block lengths and throughput made the use of FPGAs impractical. To the best of our knowledge, this remains the fastest known implementation of modified Toeplitz hashing.

We tested the GPU-based implementation using our own extractor alongside the \texttt{Validator} class. Instead of relying on files for input and output, we integrated our package with the GPU implementation using ZeroMQ queues, which are its preferred communication interface. During validation, we immediately observed discrepancies in approximately half of the tests involving randomized inputs, specifically those where the last bit was set to 1. This allowed us to quickly identify the root cause of the issue: the GPU implementation was ignoring the final bit of the input vector.

Our library helped us in identifying and fixing the issue. The GPU implementation is available online\footnote{\url{https://github.com/nicoboss/PrivacyAmplification}} and our validation is provided as an example in the repository\footnote{\url{https://github.com/cryptohslu/randextract/blob/main/examples/validation_gpu_modified_toeplitz_zeromq.py}}.

\subsection{Rust Toeplitz hashing}

Rust is a modern programming language that is designed to achieve both memory safety and high performance~\cite{rustbook}. Memory safety is a valuable feature for all software, but it is especially critical for software used in cryptographic protocols since many serious security bugs are caused by memory related issues~\cite{cisa_memory_safety}. To explore this further, a Rust implementation of Toeplitz hashing was developed at the Lucerne University of Applied Sciences and Arts as part of a semester student project\footnote{Source code is available on request.}. The primary goal was to evaluate the performance of a CPU-based Toeplitz hashing implementation within a memory-safe environment.

\begin{figure}
    \centering
    \includegraphics[width=\linewidth]{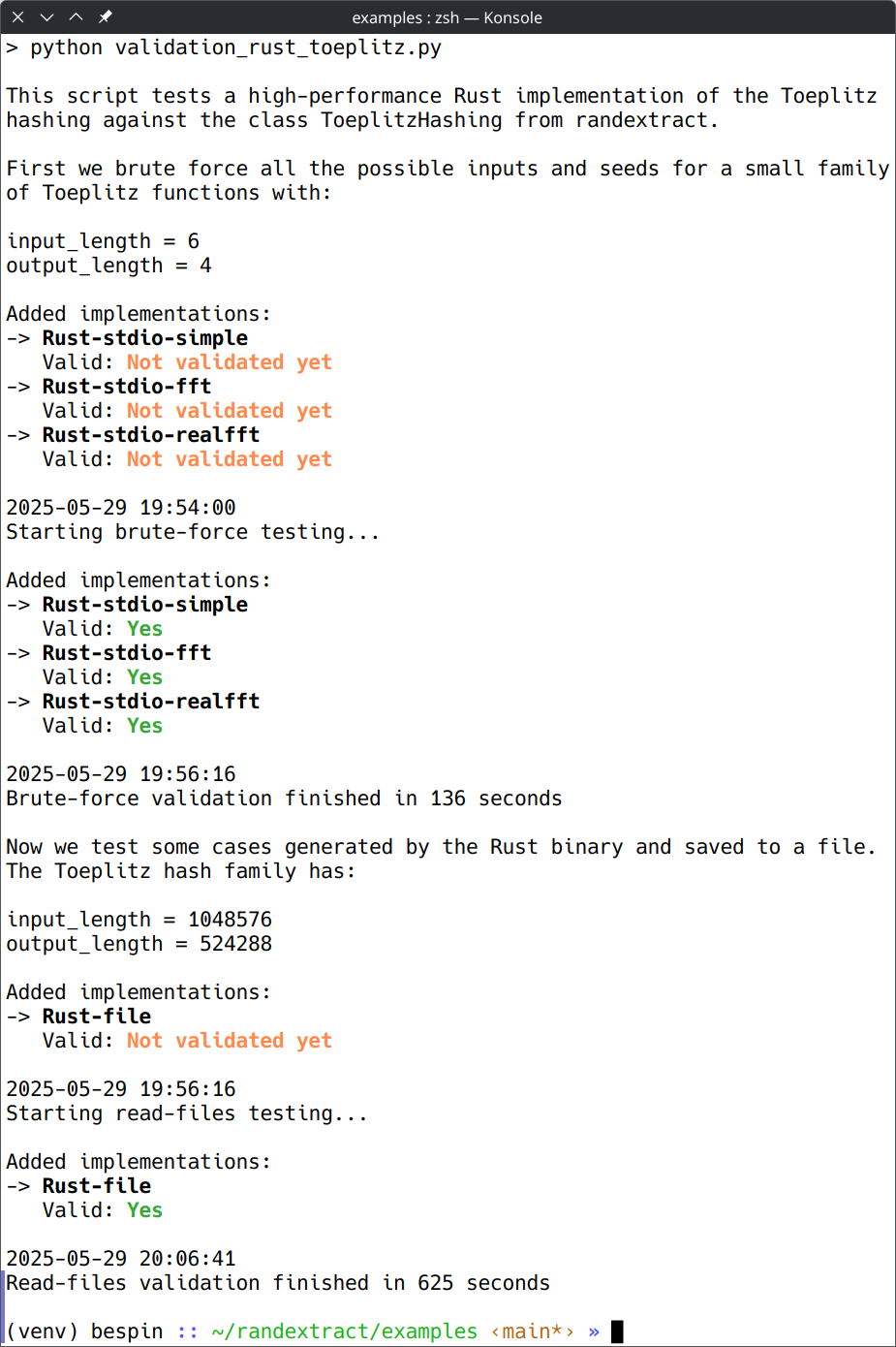}
    \caption{Screenshot showing the output of the script \lstinline|validation_rust_toeplitz.py|, available in the \texttt{examples} directory in our repository, testing the Rust implementation. First, all possible inputs and seeds were tested with very small Toeplitz matrices. Then, random samples with larger inputs and outputs were validated.}
    \label{fig:validator_rust}
\end{figure}

We used the \texttt{Validator} class to exhaustively test all possible input–seed pairs for small input lengths by interacting directly with the Rust process. For larger input vectors, we conducted randomized tests using file-based input and output. The library was employed throughout the development of the Rust implementation for continuous testing, to support identifying bugs and handling edge cases.

\subsection{C++ Trevisan's construction}

Validating Trevisan's extractors is more complex than validating extractors based on two-universal hashing. This increased complexity arises from the greater number of configurable components in Trevisan’s construction, such as the choice of one-bit extractor and weak design.

A C++ high-performance implementation of Trevisan's construction providing two different weak designs and three one-bit extractors was developed in~\cite{mauerer2012modularframeworkrandomnessextraction}. This library was used in QRNG~\cite{kavuri2024traceablerandomnumbersnonlocal} and QKD~\cite{Nadlinger2022} experiments.

When validating this implementation, we identified three issues affecting the basic weak design implementation based on finite fields. Two of these are caused by an over-optimization of the multiplication and addition operations, which are implemented using left-shift and bitwise OR operations, respectively. While such optimizations are correct for powers of two, they are generally invalid when working with finite fields of arbitrary order, such as those used in this implementation. As a result, the computed subset overlap violates the bound specified in Eq.~\eqref{eq:weak-design-bound}, indicating that the resulting family of subsets does not form a valid weak design. Consequently, the Trevisan's construction, as defined in Eq.~\eqref{eq:trevisan_construction}, using these subsets is not a strong quantum-proof extractor. The third issue concerns a deviation in how polynomial evaluation is performed. In particular, the coefficients are interpreted in reverse compared to the formulation in App.~C.1 of~\cite{mauerer2012modularframeworkrandomnessextraction}. We believe this deviation does not affect the correctness of the weak design, but it does result in outputs that differ from our implementation.

A full description of the issues and solutions are contained directly in the library repository\footnote{\url{https://github.com/wolfgangmauerer/libtrevisan/pull/2}}.

\section{\uppercase{Test vectors for Privacy Amplification}}\label{sec:test-vectors}

Classical cryptographic functions are heavily standardized. There exist standards and procedures to get certified (see e.g.,~\cite{NIST_IR_8545}). The National Institute of Standards and Technology (NIST) operates the Cryptographic Module Validation Program (CMVP), which promotes the use of validated cryptographic modules. The validation proceeds in two steps:
\begin{enumerate}
    \item First, the underlying cryptographic algorithms and their components are tested and validated through the Cryptographic Algorithm Validation Program (CAVP)~\cite{NIST-CAVP:2025}. This validation is performed using the Automated Cryptographic Validation Test System in a black-box manner. The implementation receives a list of inputs in a request file (\texttt{.req}). The implementation then obtains the outputs based on those inputs and generates a response file (\texttt{.rsp}), which contains both the provided inputs from the received request and the outputs.
    \item Then, the full cryptographic module, including software or hardware integration, is tested under the CMVP~\cite{NIST-CMVP:2025}. This step involves functional testing, documentation review, and an evaluation of the module's conformance to requirements defined in standards such as FIPS 140-3~\cite{fips140-3}. The testing is performed by accredited third-party laboratories, and successful validation results in an official NIST certificate, authorizing the module for use in regulated and security-sensitive environments.
\end{enumerate}

Other countries, such as Spain~\cite{ccn_certification} in Europe and South Korea~\cite{kcmvp} in Asia, have similar validation processes for cryptographic modules, typically based on the ISO/IEC 19790 standard~\cite{ISO_19790}.

Currently, no equivalent validation program exists for the algorithms used in quantum cryptographic devices. Most efforts have focused on the physical layer, particularly on mitigating side-channel attacks. Looking ahead, it is crucial to establish validation frameworks for the classical post-processing steps of QKD, especially for the steps that are crucial for the final security such as parameter estimation or privacy amplification, since these are the steps that guarantee the secrecy of the final key. Standardization of QKD protocols should define a set of approved PA algorithms, which would then be subject to testing in a CAVP-like program using randomly generated test vectors. These vectors should comprehensively reflect the range of inputs that the device is expected to handle in practice, taking into account both its operational capabilities and the constraints dictated by the protocol's security proof. After the successful completion of these tests, the post-processing modules can be validated under a CMVP-style certification program.

The \texttt{Validator} class in \texttt{randextract} can generate tests vectors following the same format as the request (\texttt{.req}) and response (\texttt{.rsp}) files used in CAVP.

\subsection{Example: Test Vectors for Toeplitz Hashing}

As an example, we provide below a response file for a small modified Toeplitz hashing extractor. Additional larger examples are available online\footnote{\url{https://github.com/cryptohslu/randextract/resources/test_vectors}}. Request and response files for any randomness extractor implemented in \texttt{randextract} can be generated using the \lstinline|generate_test_vector()| method from the \lstinline|Validator| class.

\begin{lstlisting}[]
# CAVS
# ModifiedToeplitzHashing
# Input Length : 128
# Compression ratio: 1/2
# Generated on Tue May 20 15:12:03 2025

[EXTRACT]

COUNT = 0
INPUT = e3fc097a6dcc77fc781a7ed3533528c8
SEED = 05f47ea39db462da99e3e29b06721ae6
OUTPUT = ab264a34f8ebc27c

COUNT = 1
INPUT = ff3d1bfe1f4c15730dc6ec1c36c7c4e8
SEED = 3eeaf730861d37e9d751d29fd6ad0ece
OUTPUT = 6411c793f97badae

COUNT = 2
INPUT = b7fa3c803d20709f25603bb1b3072917
SEED = 63296df538784e26c446211c058eb9a4
OUTPUT = 3bcfb106e23573e2

COUNT = 3
INPUT = 42c2ddcaef33a3e7998104c76605a588
SEED = 05b7c4012ffc8b5a17cdc544f3e7e2fd
OUTPUT = 48f041d38296ffcc

COUNT = 4
INPUT = a14c3632e4fbffff0e10b10ba4ccdc5d
SEED = 7733fabb766a34b3883762e240db6f20
OUTPUT = 16b0ed99752aa43a

COUNT = 5
INPUT = 23473c65a2c5ab8dbe073f8e419ccee7
SEED = 0c50697d5a102b6ef9016e809fb6a515
OUTPUT = f0b5b4d1f7cb519f

COUNT = 6
INPUT = 5b2719a61b8f72e208587b4ad0ec8ac0
SEED = 0ee322c8bfa4a7e901b3e0bcb0f8bad3
OUTPUT = b1731fb59a4bdb98

COUNT = 7
INPUT = 82c6f364c42caa101fb70e562585fc86
SEED = 29aa29456ea804ca102737d1d150e221
OUTPUT = d35034bccd12b0c4
\end{lstlisting}

\section{\uppercase{Conclusions \& Future work}}\label{sec:conclusion}

The correctness of classical post-processing, and in particular the privacy amplification step, is essential for the security of real quantum cryptographic protocols. Only through careful and rigorous validation can the keys obtained in QKD protocols be secure in practice. Deviations from correct privacy amplification procedures may enable practical attacks, allowing adversaries, even those without quantum capabilities, to partially or fully recover the secret keys. 

Our main contribution is an open-source Python library that implements quantum-proof strong extractors used in QKD and QRNG protocols. The library is designed for readability and auditability, with implementations closely following the mathematical definitions. We used this library to test, validate, and fix issues in external high-performance implementations.

Future work can add new classes of functions to \texttt{randextract} such as quantum-proof two-source (seedless) extractors. The modular design of the library will facilitate this task. Additionally, \texttt{randextract} can provide an open platform to test and validate quantum cryptographic protocols in ongoing post-processing standardization efforts at ETSI, ITU-T, etc.

\section*{\uppercase{Acknowledgments}}
This work was supported by the Swiss National Science Foundation Practice-to-Science Grant No 199084.

\bibliographystyle{apalike}
{\small
\bibliography{references}}

\begin{thebibliography}{}

\bibitem[Alagic et~al., 2025]{NIST_IR_8545}
Alagic, G., Bros, M., Ciadoux, P., Cooper, D., Dang, Q., Dang, T., Kelsey, J., Lichtinger, J., Liu, Y.-K., Miller, C., Moody, D., Peralta, R., Perlner, R., Robinson, A., Silberg, H., Smith-Tone, D., and Waller, N. (2025).
\newblock Status report on the fourth round of the nist post-quantum cryptography standardization process.
\newblock Technical Report NIST IR 8545, National Institute of Standards and Technology.
\newblock Accessed May 30, 2025.

\bibitem[Avanzi et~al., 2020]{kyber}
Avanzi, R., Bos, J., Ducas, L., Kiltz, E., Lepoint, T., Lyubashevsky, V., Schanck, J.~M., Schwabe, P., Seiler, G., and Stehlé, D. (2020).
\newblock {CRYSTALS-Kyber}: Algorithm specifications and supporting documentation.
\newblock \url{https://csrc.nist.gov/Projects/post-quantum-cryptography/post-quantum-cryptography-standardization/round-3-submissions}.
\newblock Third-round submission to the NIST Post-Quantum Cryptography Standardization Process.

\bibitem[Backes et~al., 2003]{bpw}
Backes, M., Pfitzmann, B., and Waidner, M. (2003).
\newblock A composable cryptographic library with nested operations.
\newblock In {\em CCS'03: Proceedings of the ACM Conference on Computer and Communications Security}, pages 220--230.

\bibitem[Bennett et~al., 1995]{privacy-amplification}
Bennett, C., Brassard, G., Crepeau, C., and Maurer, U. (1995).
\newblock Generalized privacy amplification.
\newblock {\em IEEE Transactions on Information Theory}, 41(6):1915--1923.

\bibitem[Bennett and Brassard, 1984]{bb84}
Bennett, C.~H. and Brassard, G. (1984).
\newblock {Quantum cryptography: Public key distribution and coin tossing}.
\newblock In {\em Proceedings of IEEE International Conference on Computers, Systems, and Signal Processing}, page 175, India.

\bibitem[Boole, 1847]{Boole1847}
Boole, G. (1847).
\newblock {\em The Mathematical Analysis of Logic: Being an Essay Towards a Calculus of Deductive Reasoning}.
\newblock Macmillan, Cambridge, UK.
\newblock Reprinted by Philosophical Library.

\bibitem[Born, 1955]{Born1955}
Born, M. (1955).
\newblock Statistical interpretation of quantum mechanics.
\newblock {\em Science}, 122(3172):675--679.

\bibitem[Bosshard et~al., 2021]{Bosshard2021}
Bosshard, N., Christen, R., H{\"a}nggi, E., and Hofstetter, J. (2021).
\newblock Fast privacy amplification on gpus.
\newblock Poster presentation at the 24th Annual Conference on Quantum Information Processing (QIP 2021), Online.

\bibitem[Canetti, 2001]{canetti}
Canetti, R. (2001).
\newblock Universally composable security: a new paradigm for cryptographic protocols.
\newblock In {\em FOCS~'01: Proceedings of the Symposium on Foundations of Computer Science}, pages 136--145.

\bibitem[Carter and Wegman, 1979]{CarterWegman1979}
Carter, J. and Wegman, M.~N. (1979).
\newblock Universal classes of hash functions.
\newblock {\em Journal of Computer and System Sciences}, 18(2):143--154.

\bibitem[{Centro Criptológico Nacional}, 2020]{ccn_certification}
{Centro Criptológico Nacional} (2020).
\newblock Esquema de evaluación y certificación de la seguridad de las tecnologías de información.
\newblock \url{https://oc.ccn.cni.es/documentos/normativa-y-legislacion/51-po-005-certificacion-de-productos-en/file}.
\newblock Accessed May 30, 2025.

\bibitem[{Cybersecurity and Infrastructure Security Agency}, 2023]{cisa_memory_safety}
{Cybersecurity and Infrastructure Security Agency} (2023).
\newblock The urgent need for memory safety in software products.
\newblock \url{https://www.cisa.gov/news-events/news/urgent-need-memory-safety-software-products}.
\newblock Accessed May 30, 2025.

\bibitem[De et~al., 2012]{deportmannvidickrenner}
De, A., Portmann, C., Vidick, T., and Renner, R. (2012).
\newblock Trevisan's extractor in the presence of quantum side information.
\newblock {\em SIAM Journal on Computing}, 41(4):915--940.

\bibitem[Diffie and Hellman, 1976]{dh}
Diffie, W. and Hellman, M.~E. (1976).
\newblock New directions in cryptography.
\newblock {\em IEEE Transactions on Information Theory}, 22(6):644--654.

\bibitem[Ekert, 1991]{ekert}
Ekert, A.~K. (1991).
\newblock Quantum cryptography based on {B}ell's theorem.
\newblock {\em Physical Review Letters}, 67(6):661--663.

\bibitem[{ETSI ISG-QKD}, 2016]{ETSI-GS-QKD011:2016}
{ETSI ISG-QKD} (2016).
\newblock {ETSI GS QKD 011 V1.1.1 — Quantum Key Distribution (QKD); Component characterization: characterizing optical components for QKD systems}.
\newblock Group Specification GS QKD 011 V1.1.1, European Telecommunications Standards Institute, Sophia Antipolis, France.

\bibitem[Gr{\"u}nenfelder, 2022]{fadriphd}
Gr{\"u}nenfelder, F. (2022).
\newblock {\em Performance, Security and Network Integration of Simplified BB84 Quantum Key Distribution}.
\newblock PhD thesis, Universit\'e de Gen\`eve.
\newblock Available at \url{https://archive-ouverte.unige.ch/unige:164897}.

\bibitem[Gr\"unenfelder et~al., 2023]{nphotonics}
Gr\"unenfelder, F., Boaron, A., Resta, G.~V., Perrenoud, M., Rusca, D., Barreiro, C., Houlmann, R., Sax, R., Stasi, L., El-Khoury, S., H\"anggi, E., Bosshard, N., Bussi\`eres, F., and Zbinden, H. (2023).
\newblock Fast single-photon detectors and real-time key distillation enable high secret-key-rate quantum key distribution systems.
\newblock {\em Nature Photonics}, 17(5):422--426.

\bibitem[Harris et~al., 2020]{Harris2020}
Harris, C.~R., Millman, K.~J., van~der Walt, S.~J., Gommers, R., Virtanen, P., Cournapeau, D., Wieser, E., Taylor, J., Berg, S., Smith, N.~J., Kern, R., Picus, M., Hoyer, S., van Kerkwijk, M.~H., Brett, M., Haldane, A., del Río, J.~F., Wiebe, M., Peterson, P., Gérard-Marchant, P., Sheppard, K., Reddy, T., Weckesser, W., Abbasi, H., Gohlke, C., and Oliphant, T.~E. (2020).
\newblock Array programming with numpy.
\newblock {\em Nature}, 585(7825):357–362.

\bibitem[Hayashi and Tsurumaru, 2016]{Hayashi2016}
Hayashi, M. and Tsurumaru, T. (2016).
\newblock More efficient privacy amplification with less random seeds via dual universal hash function.
\newblock {\em IEEE Transactions on Information Theory}, 62(4):2213--2232.

\bibitem[Hostetter, 2020]{Hostetter2020}
Hostetter, M. (2020).
\newblock {Galois: A performant NumPy extension for Galois fields}.
\newblock \url{https://github.com/mhostetter/galois}.

\bibitem[{International Organization for Standardization and International Electrotechnical Commission}, 2025]{ISO_19790}
{International Organization for Standardization and International Electrotechnical Commission} (2025).
\newblock {ISO/IEC 19790:2025: Information technology — Security techniques — Security requirements for cryptographic modules}.
\newblock \url{https://www.iso.org/standard/52906.html}.
\newblock Accessed May 30, 2025.

\bibitem[{ISO/IEC JTC 1/SC 27}, 2023]{ISO23837-2:2023}
{ISO/IEC JTC 1/SC 27} (2023).
\newblock {ISO/IEC 23837-2:2023 — Information security — Security requirements, test and evaluation methods for quantum key distribution — Part 2: Evaluation and testing methods}.
\newblock International Standard ISO/IEC 23837-2:2023, International Organization for Standardization and International Electrotechnical Commission, Geneva.
\newblock Published September 2023.

\bibitem[Kavuri et~al., 2024]{kavuri2024traceablerandomnumbersnonlocal}
Kavuri, G.~A., Palfree, J., Reddy, D.~V., Zhang, Y., Bienfang, J.~C., Mazurek, M.~D., Alhejji, M.~A., Siddiqui, A.~U., Cavanagh, J.~M., Dalal, A., Abellán, C., Amaya, W., Mitchell, M.~W., Stange, K.~E., Beale, P.~D., Brandão, L. T. A.~N., Booth, H., Peralta, R., Nam, S.~W., Mirin, R.~P., Stevens, M.~J., Knill, E., and Shalm, L.~K. (2024).
\newblock Traceable random numbers from a nonlocal quantum advantage.

\bibitem[Klabnik and Nichols, 2019]{rustbook}
Klabnik, S. and Nichols, C. (2019).
\newblock {\em The Rust Programming Language}.
\newblock No Starch Press, 2 edition.

\bibitem[K{\"o}nig and Renner, 2011]{KonigRenner2011}
K{\"o}nig, R. and Renner, R. (2011).
\newblock Sampling of min-entropy relative to quantum knowledge.
\newblock {\em IEEE Transactions on Information Theory}, 57(7):4760--4787.

\bibitem[K{\"o}nig and Terhal, 2008]{koenigterhal}
K{\"o}nig, R.~T. and Terhal, B.~M. (2008).
\newblock The bounded-storage model in the presence of a quantum adversary.
\newblock {\em IEEE Transactions on Information Theory}, 54(2):749--762.

\bibitem[Krawczyk, 1995]{Krawczyk1995}
Krawczyk, H. (1995).
\newblock New hash functions for message authentication.
\newblock In Guillou, L.~C. and Quisquater, J.-J., editors, {\em Advances in Cryptology --- EUROCRYPT '95}, pages 301--310, Berlin, Heidelberg. Springer Berlin Heidelberg.

\bibitem[Li et~al., 2019]{ToeplitzFPGA}
Li, Q., Yan, B.-Z., Mao, H.-K., Xue, X.-F., Han, Q., and Guo, H. (2019).
\newblock High-speed and adaptive fpga-based privacy amplification in quantum key distribution.
\newblock {\em IEEE Access}, 7:21482--21490.

\bibitem[Mauerer et~al., 2012]{mauerer2012modularframeworkrandomnessextraction}
Mauerer, W., Portmann, C., and Scholz, V.~B. (2012).
\newblock A modular framework for randomness extraction based on trevisan's construction.

\bibitem[Maurer, 1999]{Maurer99}
Maurer, U. (1999).
\newblock Information-theoretic cryptography.
\newblock In Wiener, M., editor, {\em Advances in Cryptology --- CRYPTO~'99}, volume 1666 of {\em Lecture Notes in Computer Science}, pages 47--64. Springer-Verlag.

\bibitem[Maurer, 2002]{Maurer02}
Maurer, U. (2002).
\newblock Indistinguishability of random systems.
\newblock In {\em EUROCRYPT '02: Proceedings of the International Conference on the Theory and Applications of Cryptographic Techniques}, pages 110--132.

\bibitem[Mendez~Veiga and Hänggi, 2025]{randextract-pypi}
Mendez~Veiga, I. and Hänggi, E. (2025).
\newblock {randextract}: A python reference implementation for testing and validating privacy amplification algorithms.
\newblock \url{https://pypi.org/project/randextract/}.
\newblock Version 0.2.0.

\bibitem[Nadlinger et~al., 2022]{Nadlinger2022}
Nadlinger, D.~P., Drmota, P., Nichol, B.~C., Araneda, G., Main, D., Srinivas, R., Lucas, D.~M., Ballance, C.~J., Ivanov, K., Tan, E. Y.-Z., Sekatski, P., Urbanke, R.~L., Renner, R., Sangouard, N., and Bancal, J.-D. (2022).
\newblock Experimental quantum key distribution certified by bell’s theorem.
\newblock {\em Nature}, 607(7920):682–686.

\bibitem[{National Institute of Standards and Technology}, 2019]{fips140-3}
{National Institute of Standards and Technology} (2019).
\newblock Security requirements for cryptographic modules.
\newblock Technical Report FIPS PUB 140-3, U.S. Department of Commerce.
\newblock Accessed May 30, 2025.

\bibitem[{National Institute of Standards and Technology}, 2024]{mlkem}
{National Institute of Standards and Technology} (2024).
\newblock {Module-Lattice-Based Key-Encapsulation Mechanism Standard}.
\newblock {Federal Information Processing Standards Publication (FIPS) NIST FIPS 203}, {Department of Commerce, Washington, D.C.}

\bibitem[{National Intelligence Service, Republic of Korea}, 2015]{kcmvp}
{National Intelligence Service, Republic of Korea} (2015).
\newblock Korean cryptographic module validation program (kcmvp).
\newblock \url{https://eng.nis.go.kr/EAF/1_7_2_1.do}.
\newblock Accessed May 30, 2025.

\bibitem[Nisan and Wigderson, 1994]{Nisan1994}
Nisan, N. and Wigderson, A. (1994).
\newblock Hardness vs randomness.
\newblock {\em Journal of Computer and System Sciences}, 49(2):149--167.

\bibitem[{NIST CAVP}, 2025]{NIST-CAVP:2025}
{NIST CAVP} (2025).
\newblock {Cryptographic Algorithm Validation Program (CAVP)}.
\newblock Program Overview CAVP, National Institute of Standards and Technology.
\newblock Programme website, accessed 30 May 2025.

\bibitem[{NIST CMVP}, 2025]{NIST-CMVP:2025}
{NIST CMVP} (2025).
\newblock {Cryptographic Module Validation Program (CMVP)}.
\newblock Program Overview CMVP, National Institute of Standards and Technology.
\newblock Programme website, accessed 30 May 2025.

\bibitem[Peterson and Weldon, 1972]{PetersonWeldon1972}
Peterson, W.~W. and Weldon, Jr., E.~J. (1972).
\newblock {\em Error-Correcting Codes}.
\newblock MIT Press, Cambridge, MA, second edition.

\bibitem[Pfitzmann and Waidner, 2001]{pw}
Pfitzmann, B. and Waidner, M. (2001).
\newblock A model for asynchronous reactive systems and its application to secure message transmission.
\newblock In {\em SP '01: Proceedings of the 2001 IEEE Symposium on Security and Privacy}, page 184.

\bibitem[{Python Software Foundation}, 2024]{python-sl}
{Python Software Foundation} (2024).
\newblock {Python Language Reference, version 3.x}.
\newblock \url{https://docs.python.org/3/}.
\newblock Accessed May 30, 2025.

\bibitem[Raz et~al., 2002]{Raz2002}
Raz, R., Reingold, O., and Vadhan, S. (2002).
\newblock Extracting all the randomness and reducing the error in trevisan's extractors.
\newblock {\em Journal of Computer and System Sciences}, 65(1):97--128.

\bibitem[Renner, 2005]{rennerphd}
Renner, R. (2005).
\newblock {\em Security of Quantum Key Distribution}.
\newblock PhD thesis, ETH Zurich.
\newblock Available at https://arxiv.org/abs/quant-ph/0512258.

\bibitem[Renner and K{\"o}nig, 2005]{rennerkoenig}
Renner, R. and K{\"o}nig, R. (2005).
\newblock Universally composable privacy amplification against quantum adversaries.
\newblock In {\em TCC'05: Proceedings of the Theory of Cryptography Conference}, pages 407--425.

\bibitem[Rivest et~al., 1983]{rsa}
Rivest, R.~L., Shamir, A., and Adleman, L.~M. (1983).
\newblock A method for obtaining digital signatures and public-key cryptosystems.
\newblock {\em Communications of the ACM}, 26(1):96--99.

\bibitem[Shaltiel, 2004]{SHALTIEL2004}
Shaltiel, R. (2004).
\newblock {\em Recent developments in explicit constructions of extractors}, page 189–228.
\newblock World scientific.

\bibitem[Stefanov et~al., 2000]{oqrng}
Stefanov, A., Gisin, N., Guinnard, O., Guinnard, L., and Zbinden, H. (2000).
\newblock Optical quantum random number generator.
\newblock {\em Journal of Modern Optics}, 47(4):595--598.

\bibitem[Tomamichel et~al., 2011]{qlhl}
Tomamichel, M., Schaffner, C., Smith, A., and Renner, R. (2011).
\newblock Leftover hashing against quantum side information.
\newblock {\em IEEE Trans. Inform. Theory}, 57(8):5524--5535.

\bibitem[Trevisan, 2001]{trevisan}
Trevisan, L. (2001).
\newblock Extractors and pseudorandom generators.
\newblock {\em J. ACM}, 48(4):860–879.

\bibitem[Virtanen et~al., 2020]{Virtanen2020}
Virtanen, P., Gommers, R., Oliphant, T.~E., Haberland, M., Reddy, T., Cournapeau, D., Burovski, E., Peterson, P., Weckesser, W., Bright, J., van~der Walt, S.~J., Brett, M., Wilson, J., Millman, K.~J., Mayorov, N., Nelson, A. R.~J., Jones, E., Kern, R., Larson, E., Carey, C.~J., Polat, I., Feng, Y., Moore, E.~W., VanderPlas, J., Laxalde, D., Perktold, J., Cimrman, R., Henriksen, I., Quintero, E.~A., Harris, C.~R., Archibald, A.~M., Ribeiro, A.~H., Pedregosa, F., van Mulbregt, P., Vijaykumar, A., Bardelli, A.~P., Rothberg, A., Hilboll, A., Kloeckner, A., Scopatz, A., Lee, A., Rokem, A., Woods, C.~N., Fulton, C., Masson, C., H\"{a}ggstr\"{o}m, C., Fitzgerald, C., Nicholson, D.~A., Hagen, D.~R., Pasechnik, D.~V., Olivetti, E., Martin, E., Wieser, E., Silva, F., Lenders, F., Wilhelm, F., Young, G., Price, G.~A., Ingold, G.-L., Allen, G.~E., Lee, G.~R., Audren, H., Probst, I., Dietrich, J.~P., Silterra, J., Webber, J.~T., Slavič, J., Nothman, J., Buchner, J., Kulick, J., Sch\"{o}nberger, J.~L., de~Miranda~Cardoso,
  J.~V., Reimer, J., Harrington, J., Rodríguez, J. L.~C., Nunez-Iglesias, J., Kuczynski, J., Tritz, K., Thoma, M., Newville, M., K\"{u}mmerer, M., Bolingbroke, M., Tartre, M., Pak, M., Smith, N.~J., Nowaczyk, N., Shebanov, N., Pavlyk, O., Brodtkorb, P.~A., Lee, P., McGibbon, R.~T., Feldbauer, R., Lewis, S., Tygier, S., Sievert, S., Vigna, S., Peterson, S., More, S., Pudlik, T., Oshima, T., Pingel, T.~J., Robitaille, T.~P., Spura, T., Jones, T.~R., Cera, T., Leslie, T., Zito, T., Krauss, T., Upadhyay, U., Halchenko, Y.~O., and Vázquez-Baeza, Y. (2020).
\newblock Scipy 1.0: fundamental algorithms for scientific computing in python.
\newblock {\em Nature Methods}, 17(3):261–272.

\bibitem[Wegman and Carter, 1981]{WegmanCarter1981}
Wegman, M.~N. and Carter, J. (1981).
\newblock New hash functions and their use in authentication and set equality.
\newblock {\em Journal of Computer and System Sciences}, 22(3):265--279.

\bibitem[Yuan et~al., 2018]{10mbitQKD}
Yuan, Z., Plews, A., Takahashi, R., Doi, K., Tam, W., Sharpe, A.~W., Dixon, A.~R., Lavelle, E., Dynes, J.~F., Murakami, A., Kujiraoka, M., Lucamarini, M., Tanizawa, Y., Sato, H., and Shields, A.~J. (2018).
\newblock 10-mb/s quantum key distribution.
\newblock {\em Journal of Lightwave Technology}, 36(16):3427--3433.

\end{thebibliography}

\end{document}